\documentclass[aps,prd,twocolumn,superscriptaddress,showpacs,showkeys]{revtex4-1}
\usepackage{graphicx,color}
\usepackage{rotating}
\usepackage{amssymb}
\usepackage{amsfonts}
\usepackage{amsmath}
\newcommand{\be}{\begin{equation}}
\newcommand{\ee}{\end{equation}}
\newcommand{\bea}{\begin{eqnarray}}
\newcommand{\eea}{\end{eqnarray}}

\begin{document}


\title{Generalized uncertainty principle impact onto the black holes information flux and the sparsity of Hawking radiation}


\author{Ana Alonso-Serrano}
\email[]{ana.alonso.serrano@aei.mpg.de}
\affiliation{Max Planck Institute for Gravitational Physics, Albert Einstein Institute, Am M\"{u}hlenberg 1, D-14476 Golm, Germany}
\affiliation{Institute of Theoretical Physics, Faculty of Mathematics and Physics, Charles University, 18000 Prague, Czech Republic}

\author{Mariusz P. D\c{a}browski}
\email[]{mariusz.dabrowski@usz.edu.pl}
\affiliation{Institute of Physics, University of Szczecin, Wielkopolska 15, 70-451 Szczecin, Poland}
\affiliation{National Centre for Nuclear Research, Andrzeja So{\l}tana 7, 05-400 Otwock, Poland}
\affiliation{Copernicus Center for Interdisciplinary Studies, S{\l}�awkowska 17, 31-016 Krak\'ow, Poland}

\author{Hussain Gohar}
\email[]{hussain.gohar@usz.edu.pl}
\affiliation{Institute of Physics, University of Szczecin, Wielkopolska 15, 70-451 Szczecin, Poland}
\date{\today}
\begin{abstract}
We investigate the generalized uncertainty principle (GUP) corrections to the entropy content and the information flux of black holes, as well as the corrections to the sparsity of the Hawking radiation at the late stages of evaporation. 
We find that due to these quantum gravity motivated corrections, the entropy flow per particle reduces its value on the approach to the Planck scale due to a better accuracy in counting the number of microstates. We also show that 
the radiation flow is no longer sparse when the mass of a black hole approaches Planck mass which is not the case for non-GUP calculations. 
\end{abstract}


\maketitle

\section{Introduction}

Since the 1970s it is known that it is possible to apply thermodynamic tools to black holes. The fact that black holes not only absorb but also emit radiation when quantum mechanics is taken into account was first proposed by Hawking \cite{hr1a,hr1aa}. This mechanism of black hole evaporation relies on the semiclassical approach, so applies the methods of quantum field theory. Hawking calculated the emission from black holes, being thermal (called Hawking radiation), and the associated temperature from which  the corresponding thermodynamic entropy was obtained \cite{bentropy}. The importance of this result lies in the issue that considering an initial pure state black hole, when it evaporates (thermally), the final state is a mixed state \cite{hr1b}. Such evolution implies a nonunitary process which is not allowed in the standard quantum mechanics. This implies loss of information during the evolution and is commonly called the black hole information puzzle. This problem has been one of the first points of discussion concerning black hole during decades and is still unsolved. There exist many different proposals that try to solve it. Some of them consider that we should forget about a unitary process, others, assuming unitary, try to find where the information of the process is lost and how to recover it. For more details, see Refs. \cite{hr11a,hr22,hr22b,hr3a,hr4a,epr,hr1,hr2,hr3,hr4,hr18}.

The thermodynamic entropy is associated with the lack of information. In this sense, we are interested in what is the entropy flux in the Hawking radiation and what we can learn about hidden information in the correlations that we decide not to see because of the coarse-graining we consider~\cite{Alonso-Serrano:2017poc}. Specifically, in this paper, we are interesting in how this radiation is modified when we take into account effects coming from pure quantum gravity.

Recently in \cite{ana1,ana2,ana3,ana4}, one of the authors of this paper, quantified the information budget in black hole evaporation. The starting point was the calculation of the budget of entropy for a black body, where the entropy flux is compensated by hidden information in the correlations, due to the emission process is unitary. This argument was extended for black holes, demonstrating that the assumption of unitarity can lead to a perfectly reasonable entropy/information budget. It was shown that during the evaporation process Bekenstein entropy of a black hole is adiabatically transferred into the Clausius entropy of the Hawking radiation field, holding during the whole evaporation process that the initial Bekenstein entropy is equal to the increasing Clausius entropy of the radiation field and the decreasing Bekenstein entropy. This study was done considering classical and quantum entropies and directly provides also that the total number of particles is proportional to the initial entropy of a black hole. 

Another interesting aspect to note, is the (extreme) sparsity of the Hawking radiation during the evaporation process \cite{matt1,page1,page2,page3,javad}. The average time between emission of successive Hawking quanta, is many times larger than the natural timescale set by the energies of the emitted quanta themselves. 

At the late stages of the Hawking evaporation process, one should take into account the strong effects coming from the underlying theory of quantum gravity that can modify Hawking temperature \cite{hr1a}  and Bekenstein entropy~\cite{bentropy}. These modifications expressed in terms of generalized uncertainty principle (GUP) have been investigated in the context of string theory \cite{stringa,stringb,stringc,stringd,stringe}, loop quantum gravity \cite{lqg1,lqg2}, modified dispersion relations and from black hole physics \cite{alpha1,GUP2,GUP3,bambi,GUP4,GUP5, alpha2,alphaposit,alphaposit2,ali1,ali2}. As a consequence of these modifications, black holes do not evaporate completely and are left with a remnant of order of Planck size with finite entropy. It was suggested that these Planck size remnants store information \cite{hr3,alpha2,chen,rabin} which gives a possible solution to the information puzzle.

It is possible to have two different kind of corrections of the entropy. First of them comes from the quantum corrections in counting microstates, keeping fixed the horizon area ({\it microcanonical} corrections)~\cite{alpha1}. These corrections will reduce the entropy as a consequence of the reduction of the uncertainty over the underlying microstates. The second kind of a correction is not related to the fundamental degrees of freedom, but corrects the entropy by considering thermal fluctuations on the horizon area ({\it canonical} corrections)~\cite{alphaposit2}. Contrary to the former one, in this latter case, the entropy increases, as one adds uncertainty to the system. In this paper we are interested in the quantum corrections when the black holes reach the Planck scale, so we will consider only the microcanonical corrections (that can also be understood as more fundamental). We investigate how these corrections modify the information flow and sparsity of the Hawking radiation. These corrections are expressed by the GUP and first of all they influence the Bekenstein entropy and the Hawking temperature, then extending onto other related quantities. 

In Sec. \ref{secII} we review the key results about the information content of black holes as given in Refs. (\cite{ana1,ana2,ana3,ana4}).  In Sec. \ref{secIII} we discuss the GUP corrections to the flow of information in black holes when they approach the Planck mass. In Sec. \ref{secIV} we study the GUP modifications to the frequency of emission of Hawking quanta. The summary of our results is given in Sec. \ref{secV}.

\section{Entropy of black holes per emitted quanta} 
\label{secII}

In Ref. \cite{ana1} the average entropy released during the standard thermodynamic process of burning a lump of coal in a blackbody furnace was calculated. It is useful to present this entropy in a dimensionless fashion, expressed in units of nats or bits. Henceforth, the entropy in nats is $\hat{S} = S/k_B$, and the entropy in bits is just defined as $\hat S_2=S/(k_B \ln 2)$, where $k_B$ is the Boltzmann constant and S the original (dimensionfull) entropy. The explicit result for an average entropy flow in black body radiation was obtained to be
\be
\langle\hat S_2\rangle=\frac{\pi^4}{30\zeta(3)\text{ln2}}~\text{bits/photon} \approx 3.90 ~\text{bits/photon}.
\label{lump}
\ee
The standard deviation of Eq. (\ref{lump}) is
\bea
\sigma_{\hat S_2}&=&\frac{1}{\text{ln2}}\sqrt{\frac{12\zeta(5)}{\zeta(3)}-\left(\frac{\pi^4}{30\zeta(3)}\right)^2}~\text{bits/photon} \nonumber \\ 
&\approx& 2.52~ \text{bits/photon},
\label{lumpdev}
\eea
where $\zeta(n)$ is the Riemann zeta function. 

This reasoning was then extended to the black hole evaporation process, determining the flux of entropy/information emitted in the form of radiation (assuming an exact Planck spectrum as first approximation)~\cite{ana2,ana3}. For the case of a Schwarzschild black hole with mass $M$, the Bekenstein entropy loss per emitted massless boson was found to be equal to the entropy content per photon in black body radiation. In addition, after calculating the gain of entropy in the Hawking flux, one finds that all the information emitted by a black hole is perfectly compensated by the entropy gain of the radiation, so that the information is completely transmitted from the hole to the external radiation \cite{ana2,ana3}.

One can also estimate the total number of emitted quanta in terms of the original Bekenstein entropy as being equal to \cite{ana2,ana3}
\begin{equation}
N = \frac{30 \zeta(3)}{\pi^4} \hat S_2 \approx 0.26 \; \hat S_2.
\label{non-GUP}
\end{equation}

According to these results, it was shown, semiclassically, how it was possible to describe the evaporation process of a black hole with an explicit and continuous flux of entropy (from the hole into the radiation), where each photon carries an amount of entropy that should be compensated by information hidden in correlations (assuming that the evaporation is a unitary process). 

\section{GUP corrections to the entropy of black holes} 
\label{secIII}

The calculations involving Hawking radiation are usually performed using semiclassical methods, but when the size of a system approaches the Planck scale, they cease to be valid and it is the full theory of quantum gravity which should be applied. As still there is no complete theory of quantum gravity, what one usually considers are some corrections to the classical theory coming from the quantum nature of spacetime. In our case of interest, we will investigate the corrections in the Hawking temperature and the Bekenstein entropy. These corrections were first suggested in string theory \cite{stringc,stringd} and in loop quantum gravity \cite{lqg1,lqg2}. In fact, it is possible to directly apply the generalized uncertainty principle (GUP) to describe such corrections \cite{alpha1,alpha2}. They are applied in order to accurately count the number of microstates that describe a black hole, and this translates into quantum corrections to microcanonical entropy.

The GUP modifies the Heisenberg principle, in presence of a gravitational field, at the Planck energies into~\cite{alpha1,GUP2,GUP3,GUP4,GUP5, alpha2,alphaposit,alphaposit2,rabin,ali1,ali2}
\be
\Delta x \Delta p = \hbar \left[1 + \alpha^2 (\Delta p )^2 \right],
\label{GUPb}
\ee
where $x$ is the position, $p$ the momentum, and
\be
\alpha = \alpha_0 \frac{l_{pl}}{\hbar}
\label{alpha}
\ee
with $\alpha_0$ a dimensionless constant, $l_{pl}$ the Planck length, $\hbar$ the Planck constant, and $\alpha$ a constant with the dimension of inverse momentum i.e. $kg^{-1}m^{-1}s$. The value of $\alpha_0$ has been largely discussed in the literature and it depends on the correction we want to apply to our system \cite{alpha1,alpha2,alphaposit,alphaposit2,alpha2011,alpha2018,alphaGW, alpha2018}. In the case that concerns us it has a negative value due to the microcanonical corrections we consider (as it was detailed in the introduction) and is of order unity on the theoretical basis (showing only our ignorance of the exact correction to the uncertainty principle from quantum gravity) \cite{alpha1, alpha2}. Recently, some investigations are discussing observational bounds over the value of $\alpha_0$~\cite{alpha2011,alpha2018,alphaGW, alpha2018a}. There have even been also claims that $\alpha_0$ is not a pure number, but depends on the ratio $m_p /M$ where M is the black hole mass and $m_p$ is the Planck mass \cite{alpha2018}.
The GUP corrections should be applied to a black hole evaporation process at its final stages (when the size of a black hole is of the magnitude of the Planck scale) and so  they imply a modification of the information/entropy carried by the emitted particles as it is explained below.

Following \cite{alpha2}, we can write Hawking temperature of a black hole by using the uncertainty relation $\Delta p\Delta x\approx \hbar$; and near the horizon of a black hole, the position uncertainty has a minimum value of $\Delta x=2l_p$, where for the case of a Schwarzschild black hole, $l_p=2GM/c^2$. This leads to an energy uncertainty
\be
\Delta pc\approx \frac{\hbar c}{\Delta x}=\frac{\hbar c^3}{4 GM} \approx k_B T.
\ee
After including a ``calibration factor'' of $2\pi$ we define Hawking temperature $T$ as 
\be
T=\frac{\hbar c^3}{8\pi Gk_BM}=\frac{c^2}{8\pi k_B}\frac{m_p^2}{M},
\ee
where, $m_p^2=\hbar c/G$, $G$ is the Newton's gravitational constant and $c$ is the speed of light. Similarly, using GUP, we can derive $T_{GUP}$. For this purpose, one can express $\Delta p$ in terms of $\Delta x$ from Eq. (\ref{GUPb}) and it gives
\be
\Delta p = \left( \frac{\Delta x}{2\hbar \alpha^2} \right) \mp \frac{\Delta x}{2\hbar \alpha^2} \sqrt{1 - \frac{4\hbar^2 \alpha^2}{(\Delta x)^2}} ,
\label{deltap}
\ee
which after expanding in series and choosing the minus sign, gives
\be
\Delta p \geq \frac{\hbar}{\Delta x} \left[ 1 + \frac{\hbar^2 \alpha^2}{(\Delta x)^2} + 2\frac{\hbar^4 \alpha^4}{(\Delta x)^4} + \ldots \right].
\label{seriesdeltap}
\ee
Again by the same argument above and using $\Delta x=2l_p=4GM/c^2$ and including the calibration factor into each term, we get 
\begin{equation}
T_{GUP} = T \left[ 1 + \frac{4\alpha^2 \pi^2  k_B^2}{c^2} T^2  + 2\left(\frac{4 \alpha^2\pi^2  k_B^2}{c^2} \right)^2 T^4 + \ldots \right].
\label{TGUP}
\end{equation}
Using this modified Hawking temperature, $T_{GUP}$, we obtain a generalized formula for Bekenstein entropy that takes into account the GUP corrections. For this purpose, we use $dS_{GUP}=c^2dM/T_{GUP}$, which gives
\begin{equation}
S_{GUP}= S - \frac{\alpha^2c^2 m_p^2 k_B   \pi}{4}\ln{\frac{S}{S_0}} + \frac{\alpha^4c^4 m_p^4k_B^2   \pi^2}{4} \frac{1}{S} + \ldots,
\label{SGUP}
\end{equation}
where $S$ is the Bekenstein entropy for a Schwarzschild black hole:
\be
S= \frac{A}{4} \frac{k_B c^3}{\hbar G} = \frac{4\pi k_BGM^2}{\hbar c}=4\pi k_B\left(\frac{M}{m_p}\right)^2 = k_B \hat{S},
\ee
and we have assumed that the integration constant $S_0 = (A_0 c^3 k_B)/{4 \hbar G}$ ($A_0=$ const. with the unit of area) in order to keep logarithmic term dimensionless.
Having these expressions, we are now able to calculate the GUP modified Bekenstein entropy loss of a black hole (following the procedure developed in Ref. \cite{ana2}), so
\bea
&& \frac{dS_{GUP}}{dN}=\frac{dS/dt}{dN/dt} \times \\
&&\left(1 -\frac{ \alpha^2c^2m_p^2 k_B  \pi}{4}\frac{1}{S} - \frac{\alpha^4c^4 m_p^4k_B^2   \pi^2}{4} \frac{1}{S^2} + \ldots \right), \label{sgup} \nonumber
 \eea
where $N$ is the number of particles and 
\be
 \frac{dS}{dN}=\frac{dS/dt}{dN/dt}=\frac{8\pi k_B}{c^2}\frac{M}{m_p^2}\hbar \langle\omega\rangle.
 \label{dSdN}
\ee
Note that we have used the mass element
\be
 dM= \frac{\langle E \rangle }{c^2} dN = \frac{\hbar \langle \omega\rangle}{c^2}dN, 
\ee
where the average energy is given by
\be
 \langle E\rangle=\hbar\langle \omega\rangle= \frac{\pi^4k_B}{30\zeta(3)} T_{GUP}. 
 \label{energy}
\ee
From now on we will use only the first two terms in Eq. (\ref{SGUP}) and Eq. (\ref{TGUP}), disregarding higher terms of $\alpha$, as the first approximation.  
This implies that the modified temperature corrects Eq. (\ref{dSdN}) in the following way
\be
\frac{dS}{dN}=\frac{k_B\pi^4}{30\zeta (3)}\left[1+\left(\frac{\alpha c}{4}\right)^2\left(\frac{m_p^2}{M}\right)^2+...\right].
\ee
Finally, calculating the terms in Eq. (\ref{sgup}), the complete modified GUP entropy (at first order in $\alpha$) will be given by 
\be
\frac{dS_{GUP}}{dN}=\frac{k_B\pi^4}{30\zeta (3)}\left[1-\left(\frac{\alpha c}{4}\right)^4\left(\frac{m_p^2}{M}\right)^4+...\right].
\label{dSGUPdN}
\ee 
\begin{figure}[h!]
\includegraphics[width=8.5cm]{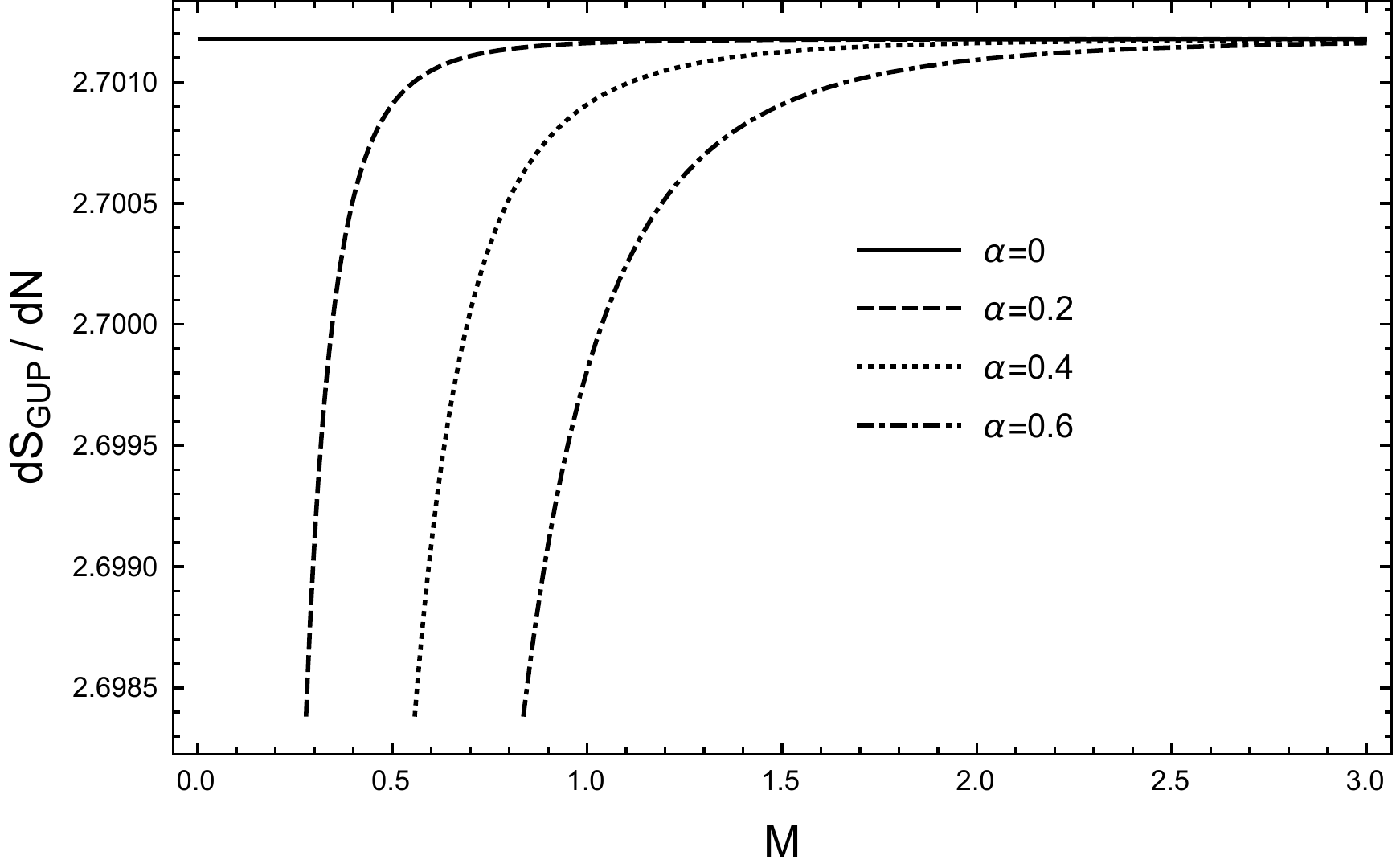}
\caption{The GUP modified Bekenstein entropy loss, $dS_{GUP}/dN,$ as given by (\ref{dSGUPdN}) as the function of $M$ for different values of $\alpha$. We have taken natural units $c$ = $G$ = $\hbar$ =1.}
\label{fig1}
\end{figure}
Now, it is possible to analyse how the entropy carried by the particles is modified when the mass approaches the Planck mass, that is, when we have strong quantum modifications. In Fig. \ref{fig1} one can see that the entropy carried per escaping particle it is not always the same (and consequently the information is not escaping at the same rate), but it reduces its value when the system is decreasing its size on the approach to the Planck length. This result makes sense together with the fact that some more accurate determination of the degrees of freedom (more information) in the system, due to the quantum correction we have introduced, reduces its entropy.
By using the GUP generalized temperature (\ref{TGUP}) we also can obtain the number of particles per emitted mass as  
\be
\frac{dN_{GUP}}{dM}=\frac{30 c^2 \zeta (3)}{\pi^4 k_BT_{GUP}}=\frac{30 c^2 \zeta(3)}{\pi^4 k_B T} \left( 1 +\frac{4 \pi^2 \alpha^2 k_B^2}{c^2} T^2 \right)^{-1},
\ee
which can be integrated to give the total number of emitted Hawking quanta $N_{GUP}$ from a black hole, when GUP corrections are present as
\be
N_{GUP}=\frac{30\zeta (3)}{\pi^4}\left[\frac{4\pi}{m_p^2} M^2  -\frac{\alpha^2 c^2m^2_p\pi}{4}\text{ln}{\left( \frac{M^2}{M_0^2}\right)}\right], 
\label{NM}
\ee
where $M$ is the initial mass of a black hole and $M_0 = (A_0 c^4)/(16 \pi G)$ is an integration constant.
\be
N_{GUP}=\frac{30\zeta (3)}{\pi^4}\left[\hat S  -\frac{\alpha^2 c^2m^2_p\pi}{4}\text{ln}{\left( \frac{\hat S }{\hat S_0}\right)}\right], \label{ngup}
\ee
in terms of entropy.This expression, compared with the semiclassical calculation of the total emitted particles [Eq. (\ref{non-GUP})], shows the introduction of a GUP modification that results in  decreasing the total of emitted particles. Such a correction makes sense considering that the final state of evaporation is a remnant of the Planck size.

\section{Sparsity of Hawking radiation modified by GUP} 
\label{secIV}

An interesting characteristic of the Hawking flux is that it is very sparse during the whole evaporation process. In order to probe this, one can use several dimensionless quantities that gave the ratio between an average time between the emission of two consecutive quanta and the natural time scale~\cite{matt1}.  

In the first approximation one assumes the exact Planck spectrum and it results in a general expression for the Minkowski spacetime, that should be specified depending on a dimensionless parameter $\eta$ (for a detailed discussion, see Ref. \cite{matt1}) given by
\begin{equation}
\eta =C\frac{\lambda_{\text{thermal}}^2}{gA},
\end{equation}
where the constant $C$ is dimensionless and depends on the specific parameter ($\eta$) we are choosing, $g$ is the spin degeneracy factor, $A$ is the area and 
\be 
\lambda_{\text{thermal}}=2\pi\hbar c/(k_BT)
\ee
 is the ``thermal wavelength.''

For a Schwarzschild black hole the temperature in the thermal wavelength is given by the Hawking temperature and the area should be replaced by an effective area (that corresponds to the universal cross section at high frequencies) given by $A_{\text{eff}}=27/4 A$, where $A$ is a horizon area of a black hole. In this case, the relevant factor in any dimensionless parameter results in
\be
\frac{\lambda^2_{thermal}}{A_{eff}}=\frac{64\pi^3}{27}\sim 73.5...\gg  1,
\ee
for massless bosons. Consequently, any of the dimensionless parameters is much larger than unity (contrary that emitters under normal laboratory conditions, where $\eta \ll 1$). This implies that the gap between successive Hawking quanta is on average much larger than the natural timescale associated with each individual emitted quantum, so the flux is very sparse. It is interesting to note that the mass $M$ of a black hole drops out during the calculations.

What is relevant here is that both the area and the ``thermal wavelength'' are GUP-modified when the system approaches the Planck scale. This results in modifying the frequency of emitted quanta from a black hole.

We obtain that the new generalized by GUP effective area is 
\be
A_{eff}|_{GUP} =\frac{27}{4}A_{GUP}=\frac{27}{4}\left[ A- \hbar^2 \alpha^2 \pi \ln{\frac{A}{A_0} } \right]
\ee
with $A_0$ an  integration constant with the unit of area, and the GUP corrected thermal wavelength is 
\be
\lambda_{\text{thermal}}|_{GUP}=\frac{2\pi\hbar c}{k_BT_{GUP}}=\frac{2\pi\hbar c}{k_BT \left[ 1 +\frac{4 \pi^2 \alpha^2 k_B^2}{c^2} T^2\right]}.
\ee 

Finally, the GUP corrected dimensionless parameter that determines the sparsity of the flux, is given by 
\bea
&& \frac{\lambda^2_{thermal}}{A_{eff}}|_{GUP}=\frac{64\pi^3}{27} \times \\
&& \frac{M^6}{  \left[M^2-(\frac{\alpha c}{4})^2m_p^4 \ln \left(\frac{M^2}{M_0^2}\right) \right] \left[M^{2}+(\frac{\alpha c}{4})^2m_p^4\right]^{2} }, \nonumber
\label{spars}
\eea
which now depends on the mass $M$ of a black hole, and on the GUP parameter $\alpha$.

\begin{figure} 
	\includegraphics[width=8.5cm]{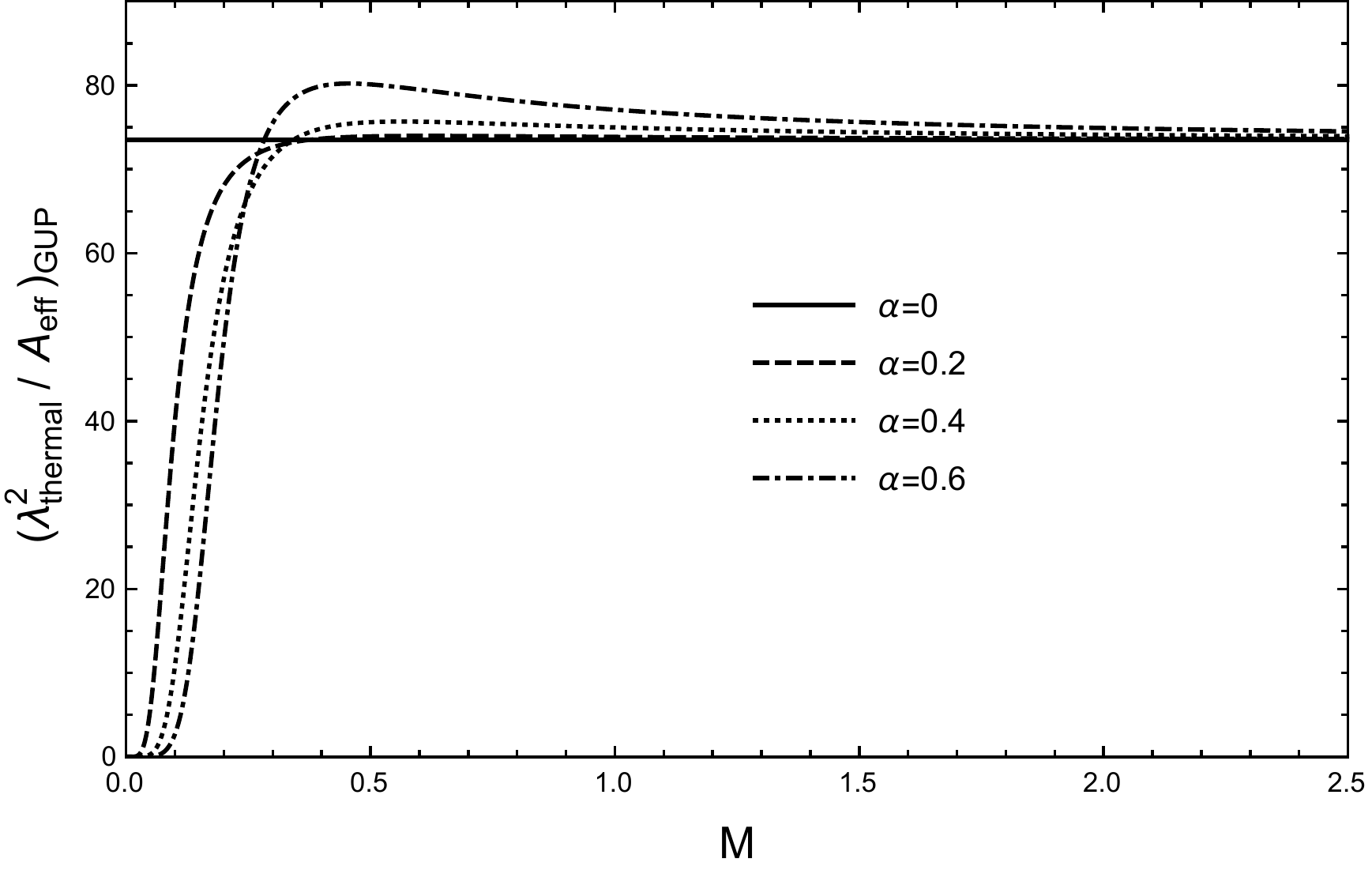}
	\caption{GUP-corrected dimensionless parameter $\frac{\lambda^2_{thermal}}{A_{eff}}|_{GUP}$ as given by (\ref{spars}) versus $M$ for different values of $\alpha$. We have taken natural units $c$ = $G$ = $\hbar$ =1}
	\label{fig4}
\end{figure}

Figure \ref{fig4} shows a plot of $\frac{\lambda^2_{thermal}}{A_{eff}}|_{GUP}$ as given by (\ref{spars}) as a function of a black hole mass for different values of $\alpha$. The horizontal line represents the case $\alpha=0$, which is consistent with the original calculation \cite{matt1}.

It is worth emphasizing again that the dimensionless parameter $\frac{\lambda^2_{thermal}}{A_{eff}}|_{GUP}$ depends on the initial mass of the black hole $M$ and the GUP parameter $\alpha$. At early stages of the Hawking evaporation process, this is not relevant (due to the huge mass of a black hole) and this dimensionless parameter is much greater than one. However, when a black hole approaches the Planck mass at the late evaporation time, the parameter approaches unity and finally becomes less than one. Note that for higher values of $\alpha$, this dimensionless parameter starts decreasing for smaller values of mass, that is for earlier phases of an evaporation process. This result indicates that the Hawking flux is no longer sparse when the black hole approaches the Planck mass, and that this regime depends of the specific GUP parameter, although the qualitative result is the same for all values of $\alpha$.
In \cite{bibhas} a similar result from calculating the sparsity of Hawking flux close to the Planck scale, using corrections coming from backreaction effects have been studied. It is interesting to note that how different approaches for that scale lead to qualitative similar results, which reassert the behavior of the Hawking flux at the final stages of evaporation.

\section{Summary}
 \label{secV}

We have investigated quantum gravity modifications to the entropy and temperature of an evaporating black hole expressed by the generalized uncertainty principle (GUP), that will be relevant when it approaches the Planck size. In this process, GUP prevents complete evaporation leaving a final remnant of the Planck size. We have shown that these modifications change the flow of information from black holes when they approach the Planck mass, and also influence amount of the sparsity of the Hawking radiation.

We have considered microcanonical corrections that are taken into account in order to more accurately count the microstates that describe a black hole.  This proper count of microstates allows to reduce the uncertainty, so the entropy is diminished in comparison to the semiclassical calculation. 

The first important result we have obtained is that the radiation flux, although continuous, does not always carry the same amount of information. Once the evaporation reaches its final stages (that is, the black hole size approaches the Planck length), the entropy budget carried by escaping particles reduces drastically, therefore the hidden information in the system also reduces. In addition is also interesting to note how also the total number of emitted particles is reduced, what is consistent with the presence of a final remnant.

The second important result is that although the Hawking radiation flux is extremely sparse during the evaporation once the semiclassical regime is considered, when we take into account the GUP modifications this is no longer the case and the radiation ceases to be sparse when the process reaches its last stages which are close to the Planck scale.

These results help to understand the behaviour of the flux of Hawking particles and the information loss from the black holes. They show how modifications coming from quantum gravity (here- the GUP corrections) change the semiclassical picture of the entropy flux, when a black holes approach last stages of their evaporation processes.

\section*{Acknowledgements}

A. A-S. is funded by the Alexander von Humboldt Foundation. A. A-S work is also supported by the Project. No. \mbox{MINECO FIS2017-86497-C2-2-P} from Spain, and partially by the grant \mbox{GACR-14-37086G} of the Czech Science Foundation. The work of M.P.D. was financed by the Polish National Science Center Grant DEC-2012/06/A/ST2/00395. The work of H.G. was financed by the Polish National Center Grant Etiuda UMO-2016/20/T/ST2/00490 and partially by the Polish National Science Center Grant DEC-2012/06/A/ST2/00395. 
This article is based upon work from COST Action CA15117 “Cosmology and Astrophysics Network for Theoretical Advances and Training Actions (CANTATA)”, supported by COST (European Cooperation in Science and Technology)


\begin{thebibliography}{99}

\bibitem{hr1a} S. W. Hawking,  Nature (London) {\bf 248}, 30 (1974).

\bibitem{hr1aa}S. W. Hawking, Commun. Math. Phys. {\bf 43}, 199 (1975).

\bibitem{bentropy} J.D. Bekenstein, Phys. Rev. D {\bf 7}, 2333 (1973).

\bibitem{hr1b} S. W. Hawking, Phys. Rev. D {\bf 14}, 2460 (1976).

\bibitem{hr11a} L. Susskind, L. Thorlacius, and J. Uglum, Phys. Rev. D {\bf 48}, 3743 (1993).

\bibitem{hr22}  G. t'Hooft, Nuc. Phys. {\bf B256}, 727 (1985). 

\bibitem{hr22b} A. Almheiri, D. Marolf, J. Polchinski, and J. Sully, J. High Energy Phys. 02, (2013) 018.

\bibitem{hr3a} A. Almheiri, D. Marolf, J. Polchinski, D. Stanford, and J. Sully, J. High Energy Phys. 09, (2013) 018.
 
\bibitem{hr4a} P.  Chen,   Y.  C.  Ong,   D.  N.  Page,   M.  Sasaki,   and D. H. Yeom, Phys. Rev. Lett. {\bf 116}. 161304 (2016)

\bibitem{epr} J. Maldacena and L. Susskind, Fortsch. Phys. {\bf 61}, 781 (2013).

\bibitem{hr18} S. W. Hawking, arXiv:1509.01147. 

\bibitem{hr1} W. G. Unruh and R. M. Wald, Rep. Prog. Phys. {\bf 80}, 092002 (2017).

\bibitem{hr2} D. Marolf, Rep. Prog. Phys. {\bf 80}, 092001 (2017).

\bibitem{hr3} S. B. Giddings, Phys. Rev. D {\bf 46}, 1347 (1992).

\bibitem{hr4} J. Polchinski, in \emph{New Frontiers in Fields and Strings: Proceedings of the 2015 Theoretical Advanced Study Institute in Elementary Particle Physics}, edited by J. Polchinski, P. Vieira and O. DeWolfe (World Scientific, Singapore, 2017), p. 353.

\bibitem{Alonso-Serrano:2017poc}
A.~Alonso-Serrano and M.~Visser, Entropy {\bf 19}, 207 (2017).
 
\bibitem{ana1} A. Alonso-Serrano and M. Visser, Phys. Lett. B  {\bf 757}, 383 (2016).

\bibitem{ana2} A. Alonso-Serrano and M. Visser, Phys. Lett. B {\bf 776}, 10 (2018).

\bibitem{ana3} A. Alonso-Serrano and M. Visser, Universe {\bf 3}, 58 (2017). 

\bibitem{ana4} A. Alonso-Serrano and M. Visser, Phys.\ Rev.\ A {\bf 96}, 052302 (2017).

\bibitem{matt1} F.~Gray, S.~Schuster, A.~Van–Brunt, and M.~Visser, Classical Quantum Gravity  {\bf 33}, 115003 (2016).

\bibitem{page1} D. N. Page, Phys. Rev. D {\bf 13}, 198 (1976).

\bibitem{page2} D. N. Page, Phys. Rev. D {\bf14 }, 3260 (1976).

\bibitem{page3} D. N. Page, Phys. Rev. D {\bf16}, 2402 (1977).

\bibitem{javad} J. T. Firouzjaee and G. F. R. Ellis, Eur. Phys. J. C {\bf76}, 620 (2016).

\bibitem{stringa} D. Amati, M. Ciafaloni, and G. Veneziano,  Phys. Lett. B {\bf 197}, 81 (1987).

\bibitem{stringb}David J. Gross and Paul F. Mende,  Phys. Lett. B {\bf 197}, 129 (1987).

\bibitem{stringc}  D. Amati, M. Ciafaloni, and G. Veneziano, Phys. Lett. B {\bf 216}, 41 (1989).

\bibitem{stringd} K. Konishi, G. Paffuti, and P. Provero, Phys. Lett. B {\bf 234}, 276 (1990).

\bibitem{stringe} A. Kempf, G. Mangano, and R. B. Mann, Phys. Rev. D {\bf 52}, 1108 (1995)
 
\bibitem{lqg1} C. Rovelli, Phys. Rev. Lett. {\bf77}, 3288 (1996).

\bibitem{lqg2} K. A. Meissner, Classical Quantum Gravity {\bf21}, 5245 (2004).

\bibitem{chen} P. Chen, Y.-Ch. Ong, D.H. Yeom,  Phys. Rep. {\bf 603}, 1 (2015). 

\bibitem{alpha2} R. J. Adler, P. Chen, and D. I. Santiago, Gen. Relativ. Gravit. {\bf 33}, 2101 (2001). 

\bibitem{rabin} R. Banerjee and S. Ghosh, Phys. Lett. B {\bf 688}, 224 (2010).

\bibitem{GUP4} A. N. Tawfik and A. M. Diab, Int. J. Mod. Phys. D {\bf 23}, 1430025 (2014).

\bibitem{GUP5} A. N. Tawfik and A. N. Diab, Int. J. Mod. Phys. A {\bf 30}, 1550059 (2015).

\bibitem{GUP2} M. Maggiore, Phys. Lett. B {\bf 304}, 65 (1993).

\bibitem{GUP3} M. Maggiore, Phys. Rev. D {\bf 49}, 5182 (1994).

\bibitem{alphaposit} A. Chatterjee and P. Majumdar,  Phys. Rev. Lett. {\bf 92}, 141301 (2004).

\bibitem{alpha1} A. J. M. Medved and E. C. Vagenas,  Phys. Rev. D {\bf 70}, 124021 (2004).

\bibitem{alphaposit2}  G. Gour and A. J. M. Medved,  Classical Quantum Gravity {\bf 20}, 3307 (2003).

\bibitem{ali1} A. Ovgun and K. Jusufi,  Eur. Phys. J. Plus {\bf131}, 177 (2016). 

\bibitem{ali2} A. Ovgun and K. Jusufi,  Eur. Phys. J. Plus {\bf132}, 298 (2017).

\bibitem{bambi} C. Bambi and F. R. Urban, Classical Quantum Gravity {\bf 25}, 095006 (2008).

\bibitem{alpha2011} A. F. Ali, S. Das, E. C. Vagenas, Phys. Rev. D {\bf84}, 044013 (2011)

\bibitem{alpha2018} E. C.Vagenas, S. M. Alsaleh, and A. F. Ali, Europhys. Lett. {\bf120}, 40001 (2017).

\bibitem{alphaGW} Z. W. Feng, S. Z. Yang b, H. L. Li, and X. T.  Zu, Phys. Lett. B {\bf 768}, 81 (2017).

\bibitem{alpha2018a} F. Scardigli, G. Lambiase, and E. C. Vagenas, Phys. Lett. B {\bf 767} , 242 (2017).

\bibitem{bibhas} A. Paul and B. R. Majhi, Int. J. Mod. Phys. A {\bf32}, 1750088 (2017).


\end{thebibliography}
\end{document}